\documentclass[preprintnumbers,amsmath,amssymb,preprint,aps]{revtex4}

\usepackage{graphicx}

\begin{document}

\title{Stochastic description of delayed systems}
\author{L. F. Lafuerza, R. Toral}

\affiliation{IFISC, Instituto de F{\'\i}sica Interdisciplinar y Sistemas Complejos, CSIC-UIB,  Campus UIB, E-07122 Palma de Mallorca, Spain  }

\date{\today}

\begin{abstract}
We study general stochastic birth and death processes including delay. We develop several approaches for the analytical treatment of these non-Markovian systems, valid, not only for constant delays, but also for stochastic delays with arbitrary probability distributions. The interplay between stochasticity and delay and, in particular, the effects of delay in the fluctuations and time correlations are discussed.
 \end{abstract}
\maketitle

%
\section{Introduction}
Stochastic modeling plays an important role in many areas of science, such as physics, ecology or chemistry \cite{VK}. Stochasticity may appear due to the lack of complete knowledge about all the relevant variables, the precise dynamics of the system or the interactions with the environment. In some cases, one can obtain a compact description of a complicated system considering only a few relevant variables but at the expense of losing deterministic predictability. Often, probabilities for some fundamental processes can be assigned on the basis of symmetries and other considerations, or on empirical analyis, and the dynamics of the process can be derived bottom-up.

Stochasticity appears together with delay terms in many situations of interest, such as gene regulation \cite{lewis,barrio,bratsun}, physiological processes \cite{Longtin90} or postural control \cite{ohira2009,Longtin10}. The combined effects of stochasticity and delay are, however, not completely understood. From the mathematical point of view, stochastic processes including delay are difficult to analyze due to the non-Markovian character. Most of the previous approaches have focused on stochastic differential equations, that consider continuous variables \cite{kuchler,Longtin99,Frank02,Frank03,ohira2000}, or random walks in discrete time \cite{ohira95,Ohira08}, where delay can be taken into account increasing the number of variables. Models with discrete variables but continuous time are the natural description of many systems such as chemical reactions, population dynamics or epidemic spreading. In some cases, discreteness can be a mayor source of fluctuations, not well captured by continuous models \cite{aparicio}. The approach with discrete variables and continuous time was used in \cite{bratsun,Galla,Mieckisz,creationdelay}. Most often, the delay time is taken to be a constant with zero fluctuations. This is not very realistic in the applications, since it is unusual to have a deterministic delay when the rest of the dynamics is stochastic. We will take this consideration into account by allowing the delay times to be random variables with arbitrary probability density functions.

In this work we study some simple, yet general, stochastic birth and death processes including delay. We will develop tree different approaches to the analytical study of this kind of non-Markovian processes, in the general case of stochastically distributed delay: a direct approach in subsection (\ref{sec:simple}), an effective Markovian reduction in subsections (\ref{sec:elaborated}) and (\ref{sec:fullfeedback}), and a master equation approach, together with a time-reversal invariance assumption, in section (\ref{dc}). The first direct approach method is interesting for its simplicity, but its application is limited to systems with first order reactions and without feedback. The second one, effective Markovian reduction, is rather flexible and general and its development is one of the main advances of this paper. The last master equation approach complements the previous, giving information about the full probability distribution. The main limitation of all the approaches is the need to assume that completion times for delayed reactions are independent random variables (independent of each other and of other variables of the system), although the initiation rates may depend on the state of the system, allowing, for example, for feedback and crowding effects, so we do not consider this limitation to be very relevant for practical applications. Although our methodology is rather general, we present it here using specific examples that have been grouped in two categories: delay in the degradation (section \ref{dd}) and delay in the creation (section \ref{dc}). We end the paper with a brief discussion and comments in section \ref{comments}. Some more technical details are left for the two appendices.

\section{Delayed degradation}
\label{dd}
We will start by studying simple stochastic birth and death processes that include delay in the degradation step. A process of this type was proposed in \cite{bratsun} as a model for protein level dynamics with a complex degradation pathway.

\subsection{Simple Case}
\label{sec:simple}
We consider first the simplest possible process including delayed degradation:
\begin{equation}
\emptyset {{C \atop \longrightarrow}\atop{}} X, \,X {{\atop \Longrightarrow}\atop{\tau}} \emptyset,
\end{equation} 
that is, a particle $X$ is created at a rate $C$ and disappears (``dies" or ``degrades") a time $\tau$ after created. We allow the delay time $\tau$ to be randomly distributed i.e. the lifetimes $\tau$ of the created particles are random variables, that for simplicity we consider independent and identically distributed, with probability density $f(\tau)$. Although not considered in this paper, the case of non-identically distributed delay times, in particular a probability density that depends on the time from birth, can also be treated. However, as commented above, the case of non-independent delay times does not seem to be tractable with the methods we present below.

We note first that distributed delay is completely equivalent to degradation at a rate that depends on the ``age" $a$ (time form creation) of the particle, i.e., processes
\begin{equation}
X {{\atop \Longrightarrow}\atop{\tau}} \emptyset, \hspace{1cm}\text{and} \hspace{1cm} \,X {{\gamma(a) \atop \longrightarrow}\atop{}} \emptyset,
\end{equation}
are equivalent if the rate $\gamma(a)$ and the probability density of the delay $f(\tau)$ are related by:
\begin{equation}
 \gamma(a)=\frac{f(a)}{\hat{F}(a)}\Rightarrow f(\tau)=\gamma(\tau)e^{-\int_{0}^{\tau}da\gamma(a)},
\end{equation}
with $\hat F(t)=1-F(t)$ being $F(t)=\textrm{Prob}(\tau<t)=\int_0^\tau d\tau f(\tau)$ the cumulative distribution of the delay-time. This is so because $\gamma(a)da$ is the probability of dying at the time interval $(a,a+da)$, if the particle is still present at $a$, and so it is nothing but the probability $f(a)da$ that the delay time $\tau$ belongs to that same interval conditioned to the particle still being alive at time $a$, an event with probability $\hat F(a)$. 
In the notation of \cite{Papoulis}, $\gamma(a)$ is nothing but the {\sl conditional failure rate}. We take $t=0$ as the time origin, so the number of alive particles at time $t$ is $n(t)=0$ for $t\le0$. Let $P(n,t)$ the probability of $n$ particles being alive at time $t$. In the remaining of this subsection we assume that there is no feedback, in the sense that the creation rate $C$ is independent on the number of particles $n$, but, for the sake of generality, we do allow it to be a function of time $C(t)$. The non-feedback assumption allows us to obtain a full analytical solution.
As shown in the appendix, independently of the form of the delay distribution, $P(n,t)$ follows a Poisson distribution
\begin{equation}
P(n,t)=e^{-\langle n(t)\rangle}\frac{\langle n(t)\rangle^n}{n!},
\end{equation}
with average $\langle n(t)\rangle=\int_{0}^tdt'C(t')\hat{F}(t-t')$. If the creation rate, $C(t)$, is independent of time, a steady state is reached, in which the average number of particles is $\langle n\rangle_\textrm{st}=C\langle\tau\rangle$, again independently of the form of the delay distribution.

\begin{figure}[h]
\centering
\includegraphics[scale=0.5,angle=0,clip]{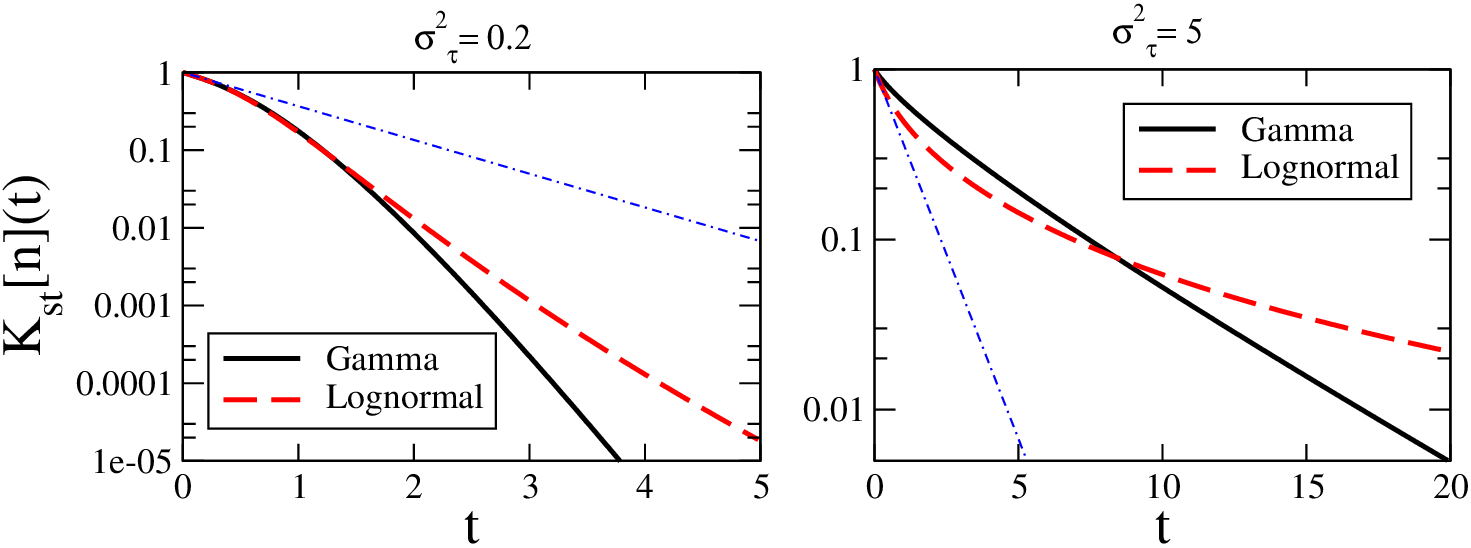}
\caption{Steady state correlation function, Eq.(\ref{corrsimple}), as a function of time, plotted in logarithmic scale, for two different types of delay distribution, gamma and lognormal, for two values of the variance of the delay: $\sigma^2_\tau=0.2$ (left panel) and  $\sigma^2_\tau=5$ (right panel); in both cases the average delay is $\langle\tau\rangle=1$ and the creation rate is $C=1$. We also plot a exponential decay with exponent one (dot-dashed line), for comparison. Note that delay distributions with larger variance and fatter tayls display slower asymptotic decay. (Online version in colour.)
 \label{correlationsimple}} 
\end{figure}

We will now compute the time correlation function. We shall see that its analytical expression does depend on the form of the delay distribution. We start from the relation:
\begin{equation}
 \langle n(t+T)|n(t)\rangle=\langle n_{new}(t+T)|n(t)\rangle+\langle n_{old}(t+T)|n(t)\rangle,
\end{equation}
with $n_{new} (n_{old})$ particles created after (before) $t$. $n_{new}$ can be computed exactly as before (now taking $t$ as the time origin), so we have:
\begin{equation}
 \langle n_{new}(t+T)|n(t)\rangle=\int_{0}^{T}dt'C(t+t')\hat{F}(T-t').
\end{equation}
The evolution of the number of particles already present at $t$ depends on the age $a$ of these particles. Their survival probability until time $t+T$ can be written as:
\begin{eqnarray}
 &&P(\text{alive at $t+T$}|\text{alive at } t)=\int_{0}^{t}daP(\text{age}=a|\text{alive at } t)P(\text{lifetime}>a+T|\text{lifetime }>a)=\nonumber\\
&&\int_{0}^{t}da\frac{C(t-a)\hat{F}(a)}{\int_{0}^tdt'C(t')\hat{F}(t-t')}\frac{\hat{F}(a+T)}{\hat{F}(a)}=\frac{\int_{0}^tdt'C(t')\hat{F}(t+T-t')}{\int_{0}^tdt'C(t')\hat{F}(t-t')},
\end{eqnarray}
where we used $P(a|b)=\frac{P(a;b)}{P(b)}$, so we find:
\begin{equation}
 \langle n_{old}(t+T)|n(t)\rangle=n(t)\frac{\int_{0}^tdt'C(t')\hat{F}(t+T-t')}{\int_{0}^tdt'C(t')\hat{F}(t-t')}.
\end{equation}
From this, one easily obtains the correlation function: $K[n](t,T)=\langle n(t)\langle n(t+T)|n(t) \rangle\rangle-\langle n(t)\rangle\langle n(t+T)\rangle$
\begin{equation}
 K[n](t,T)=\int_0^tdt'C(t')\hat{F}(t+T-t')\label{corrsimple},
\end{equation}
If $C(t)=C$, independent of time, a steady-state can be reached with correlation function $K_{\textrm{st}}[n](T)=\lim_{t\to\infty}K[n](t,T)$. For a constant rate $\gamma$, which would be equivalent to an exponential delay distribution $f(\tau)=\gamma e^{-\gamma\tau}$, it has the usual exponential decay $K_{\textrm{st}}[n](T)=(C/\gamma) e^{-\gamma T}$. For a fixed delay time $\tau_0$, corresponding to $f(\tau)=\delta(\tau-\tau_0)$, the correlation function is a straight line $K_{\textrm{st}}[n](T)=C(\tau_0-T)$ for $T<\tau_0$ and $K_{\textrm{st}}[n](T)=0$ for $T\ge \tau_0$. For other distributions of delay time, the correlation function adopts different forms, but it is always monotonically decreasing. In figure (\ref{correlationsimple}) we plot the correlation function for two different types of distribution of delay, for different values of the variance of the delay. We see that the distribution with fatter tail displays a slower asymptotic decay, and that the decay is slower as the variance of the delay increases. Numerical simulations, performed with a conveniently modified version of the Gillespie algorithm \cite{Cai}, are in perfect agreement with this exact result, providing a check of its correctness. We remark that the functional form of the decay of the correlation function depends on the delay distributed and can differ from the exponential decay found in systems without delay.

\subsection{More elaborated Case}\label{sec:elaborated}
We now consider a process including both instantaneous and delayed degradation steps:
\begin{equation}
\label{cx0}
\emptyset {{C \atop \longrightarrow}\atop{}} X, \, X {{\gamma \atop \longrightarrow}\atop{}} \emptyset, \,X{{D \atop \longrightarrow}\atop{}}{{\atop\Longrightarrow}\atop{\tau}} \emptyset,
\end{equation}
this is, particles are created at a rate $C$ and each particle can be eliminated by two processes: i) instantaneous degradation at a rate $\gamma$; ii) delayed degradation, initiated at a rate $D$ but completed only a time $\tau$ after initiation. Again, we will allow the delay-degradations times to be random variables that, for simplicity, will be independent and identically distributed with probability density function $f(\tau)$. 

For the process to be completely defined, one has to specify if a particle that initiates delayed-degradation at time $t$ and thus will disappear at $t+\tau$ (this kind of particles will be called ``infected"), can also disappear before the completion of this reaction, through instantaneous degradation. In the most general case, this can happen at a rate $\gamma'$, not necessarily equal to $\gamma$. Note that, in the case of first-order degradation ($\gamma'$ not dependent on the number of particles $n$), this instantaneous degradation is completely equivalent to a system with $\gamma'=0$, after modifying the distribution of the delayed-degradation times in the following way:
\begin{equation}
\label{ftau2}
 f(\tau)\to e^{-\gamma'\tau}f(\tau)+e^{-\gamma'\tau}\gamma'\hat F(\tau).
\end{equation}
That is, when instantaneous degradation is added to infected particles, the probability that the lifetime is equal to $\tau$ has two contributions: (i) a particle initially has a lifetime $\tau$ (probability density $f(\tau)$) and survives up to this time (an event with probability $e^{-\gamma' \tau}$); (ii) a particle has a lifetime larger than $\tau$ (probability $\hat F(\tau)$), but survives up to $\tau$ (probability $e^{-\gamma' \tau}$) and then undergoes instantaneous degradation (at rate $\gamma'$). The consideration of these two contributions leads straightforwardly to Eq.(\ref{ftau2}). We see that omitting first order instantaneous degradation of infected particles comprises no loss of generality, given that the treatment is valid for general distributions of delay. 

If $D$ and $\gamma$ are independent of $n$, the process is equivalent to the one-variable system discussed in the previous subsection \ref{sec:simple}) with a conveniently modified distribution of delay:
\begin{equation}
f(\tau)\to e^{-(\gamma+D)\tau}\gamma+\int_{0}^{\tau}dt'e^{-(\gamma+D)t'}Df(\tau-t').
\end{equation}
This comes from the fact that a particle may disapear at time $\tau$ because it did not disapeared or was infected before and is degraded instantaneously (probability density $e^{-(\gamma+D)\tau}\gamma$) or because it got infected at some previous time ($t'$) with an appropriate lifetime ($\tau-t'$, probability density $\int_{0}^{\tau}dt'e^{-(\gamma+D)t'}Df(\tau-t')$). This includes as particular cases the ones studied in \cite{Mieckisz,degradationdelay}. The results of subsection (\ref{sec:simple}) allows us to obtain the full solution also in the general case of distributed delay. If $D$ or $\gamma$ depend on $n$ the processes are not anymore equivalent, two variables are necessary and a new approach is needed for the analysis. In the following we develop this method. We will also consider the case in which the creation rate $C$ depends on the number of particles.


The full process corresponds to the following two-variable system:
\begin{equation}\label{stoproc}
\emptyset {{C \atop \longrightarrow}\atop{}} X_{A}, \,X_{A} {{\gamma \atop \longrightarrow}\atop{}} \emptyset, \,X_{A} {{D \atop \longrightarrow}\atop{}} X_{I}+Z, \,X_{I} {{\atop \Longrightarrow}\atop{\tau}} \emptyset,
\end{equation}
where we have split the proteins into two types: $X_{I}$ are infected particles that will die precisely at a time $\tau$ (itself a stochastic variable) after being infected and $X_{A}$ are non-infected (``active") particles (so $X=X_{A}\cup X_{I}$). We allow the rates to depend on $n_A$, the number of $X_A$, active, particles, but not on  $n_I$, the number of $X_I$, infected, particles which are considered to be ``inert"; this condition will be realxed in the next subsetion. Following \cite{Mieckisz}, we have introduced the auxiliary particles $Z$ whose number is given by the stochastic variable $n_Z(t)$. The introduction of $Z$ will allow us to obtain the properties of $n_I$ by using the relation:
\begin{equation}
 n_I(t)=\int_{-\infty}^{t}dt'\frac{dn_{Z}(t')}{dt'}s(t',t),\label{nonmarkov}
\end{equation}
where the discrete process $n_Z(t)$ is a sequence of step (Heaviside) functions and its derivative must be understood as a series of Dirac-delta functions. Here we have introduced the family of ``survival"  stochastic processes $s(t',t)$ defined in the following way: first, for each $t'$ we obtain a value of $\tau(t')$ independently drawn from the distribution $f(\tau)$.  Next, we set $s(t',t)=1$, if $t\in(t',t'+\tau(t'))$, and $s(t',t)=0$, otherwise. This can be considered as the indicator function of a virtual \footnote{ $s(t',t)$ is defined for all $t'$, regardless if a particle is actually infected a time $t'$. However it only contributes to (\ref{nonmarkov}) if a particle is actually infected at time $t'$, since only then $\frac{dn_{Z}(t')}{dt'}\neq0$} particle that is infected at $t'$ and survives up to a time $t'+\tau(t')$. It follows from the definition that:
\begin{eqnarray}
\langle s(t_1,t)\rangle&=&\hat F(t-t_1),\\
\langle s(t_1,t)s(t_2,t')\rangle&=&\begin{cases}\langle s(t_1,t)\rangle\langle s(t_2,t')\rangle&\text{if } t_1\ne t_2\\
\langle s(t_1,\max\{t,t'\})\rangle&\text{if } t_1=t_2,\end{cases}\label{corrsurvival}
\end{eqnarray}

Expressions (\ref{nonmarkov}-\ref{corrsurvival}) are the main advances of this section and provide us with the necessary tools to derive the main properties of the stochastic process (\ref{cx0}). In the case considered in \cite{Mieckisz} there is a fixed delay ($f(\tau)=\delta(\tau-\tau_0)$) and no instantaneous degradation of infected particles ($\gamma'=0$), so one has simply $n_I(t)=n_Z(t)-n_Z(t-\tau)$
. The inclusion of the survival process  $s(t',t)$ allows us to consider the general case of distributed delay and rates depending on the state of the system.

Note that the process followed by $\{n_A,n_Z\}$ is Markovian as the delay only appears in variable $X_I$, so the properties of $n_Z$ can be obtained using Markovian methods, and the properties of the variable $n_I$ can be derived afterwards using (\ref{nonmarkov}-\ref{corrsurvival}).
In particular, the first moments follow:
\begin{eqnarray}
 \langle n_I(t)\rangle&=&\int_{-\infty}^tdt' \frac{d\langle n_Z(t')\rangle}{dt'}\langle s(t',t)\rangle,\label{averagenz}\\
\langle n_I(t)n_I(t+T)\rangle&=&\int_{-\infty}^tdt_1\int_{-\infty}^{t+T}dt_2\frac{d^2\langle n_Z(t_1)n_Z(t_2)\rangle}{dt_1dt_2}\langle s(t_1,t)s(t_2,t+T).\rangle\label{correlationnz}
\end{eqnarray}
Using standard Markovian methods \cite{VK}, one can prove that the process $\{n_A,n_Z\}$ is described by the master equation:
\begin{eqnarray}
 \frac{dP(n_A,n_Z,t)}{dt}&=&(E^{-1}_A-1)C(n_A)P(n_A,n_Z,t)+(E_A-1)\gamma(n_A) P(n_A,n_Z,t)\nonumber\\
&+&(E_AE^{-1}_Z)D(n_A)P(n_A,n_Z,t),
\end{eqnarray}
with $E_i$ the step operator, $E_if(n_i,n_j)=f(n_i+1,n_j)$. In this section, we allow the creation rate $C$ to depend on the number of $X_A$-particles, constituting a feedback term on the number of "active" particles. From the master equation one easily derives the equations for the moments, the first of them read:
\begin{eqnarray}
 \frac{d\langle n_A\rangle}{dt}&=&\langle C(n_A)\rangle-\langle(\gamma+D)n_A\rangle\label{na},\\
\frac{d\langle n_Z\rangle}{dt}&=&\langle Dn_A\rangle\label{nz},\\
\frac{d\langle n_A^2\rangle}{dt}&=&2\langle 2(n_A+1)C(n_A)\rangle-\langle (2n_A-1)n_A(\gamma+D)\rangle),\\
\frac{d\langle n_Z^2\rangle}{dt}&=&2\langle Dn_An_Z\rangle+\langle Dn_A\rangle\\
\frac{d\langle n_An_Z\rangle}{dt}&=&\langle C(n_A)n_Z\rangle-\langle (\gamma+D)n_An_Z\rangle+\langle D(n_A^2-n_A)\rangle).
\end{eqnarray}
In the case that $C(n_A)$ is a linear function of $n_A$ and $\gamma$ and $D$ do not depend on $n_A$ (and none of them depend on $n_I$ or $n_Z$), the system of equations is closed and can be solved. For non-linear systems, we will make use of van Kampen's expansion \cite{VK}. This is a standard systematic expansion of the master equation, that consists on assuming a deterministic,  and a stochastic part for the variables that scale differently with a large parameter, $\Omega$ (typically the volume or system size) i.e. $n_A=\Omega \phi_A(t)+\Omega^{1/2}\xi_A$, $n_Z=\Omega \phi_z(t)+\Omega^{1/2}\xi_Z$. One can then write the master equation for the new variables $\xi_A$, $\xi_Z$ and expand in powers of $\Omega^{-1/2}$. The method is generically valid, provided that the rates depend on the variables only trough $n_\alpha/\Omega$ (plus higher orders in $\Omega^{-1}$; a common factor depending  on $\Omega$ multiplying all rates is also acceptable), which is fulfilled by most systems of interest, and that the macroscopic equations have a steady state as a single attractor. The equations for the macroscopic components are:
\begin{eqnarray}
 \frac{d\phi_A}{dt}&=&C(\phi_A)-[\gamma(\phi_A)+D(\phi_A)]\phi_A,\label{fia}\\
\frac{d\phi_z}{dt}&=&D(\phi_A)\phi_A.
\end{eqnarray}
The stochastic contributions, to first order in $\Omega^{-1/2}$, read:
\begin{eqnarray}
 \frac{d\langle\xi_A\rangle}{dt}&=&-\left[\widetilde{\gamma}+\widetilde{D}-C'(\phi_A)\right]\langle\xi_A\rangle\label{xia},\\
\frac{d\langle\xi_Z\rangle}{dt}&=&\widetilde{D}\langle\xi_A\rangle\label{xiz},\\
\frac{d\langle\xi_A^2\rangle}{dt}&=&-2\left[\widetilde{\gamma}+\widetilde{D}-C'(\phi_A)\right]\langle\xi_A^2\rangle+\left(\widetilde{\gamma}+\widetilde{D}\right)\phi_A+C(\phi_A),\\
\frac{d\langle\xi_Z^2\rangle}{dt}&=&2\widetilde{D}\langle\xi_A\xi_Z\rangle+\widetilde{D}\phi_A,\\
\frac{d\langle\xi_A\xi_Z\rangle}{dt}&=&-\left[\widetilde{\gamma}+\widetilde{D}-C'(\phi_A)\right]\langle\xi_A\xi_Z\rangle+\widetilde{D}(\langle\xi_A^2\rangle-\phi_A),\label{xiaxiz}
\end{eqnarray}
with $\widetilde{D}\equiv D(\phi_A)+D'(\phi_A)\phi_A$, $\widetilde{\gamma}\equiv\gamma(\phi_A)+\gamma'(\phi_A)\phi_A$. 
Usually, for the ansatz about the scaling of the variables to work (and so the expansion), the equations for the macroscopic components must have a single stable fixed point. In this case, however, the equation for $\phi_z$ does not have a fixed point, and $\phi_z(t)$ and $\langle\xi_Z^2(t)\rangle$ grow without bound. This grow, nevertheless, is consistent with $\frac{\langle n_Z(t)\rangle}{\sigma^2[n_Z](t)}=O(\Omega^0)$, and the expansion can still be applied.

(\ref{xia}-\ref{xiaxiz}) is a system of closed linear equations and so can always be solved. To compute the time correlations of $n_I$ from Eq.(\ref{correlationnz}) we need the time correlations of $n_Z$. We note that:
\begin{eqnarray}
 \langle n_Z(t_1)n_Z(t_2)\rangle&=&\sum_{n_{Z1},n_{Z2}}n_{Z1}n_{Z2}P(n_{Z2},t_2;n_{Z1},t_1)\\
 &=&\sum_{n_{Z1},n_{Z2},n_A}n_{Z1}n_{Z2}P(n_{Z2},t_2|n_{Z_1},n_A,t_1)P(n_{Z,1},n_A,t_1)\nonumber\\
&=&\left\langle \langle n_Z(t_2)|n_{Z}(t_1),n_A(t_1)\rangle n_{Z}(t_1) \right\rangle,
\end{eqnarray}
and that $\langle n_Z(t_2)|n_{Z}(t_1),n_A(t_1)\rangle$ (for $t_2>t_1$) can be obtained integrating (\ref{na}-\ref{nz}) or (\ref{xia}-\ref{xiz}).
In the general, non-linear, case, using first order van Kampen's expansion, one obtains, over the steady state:
\begin{equation}
 \langle n_Z(t_1)n_Z(t_2)\rangle=\Omega^2\phi_z(t_1)\phi_z(t_2)+\Omega\left[\langle\xi_Z^2(\min\{t_1,t_2\})\rangle+\frac{\widetilde{D}}{u}\langle\xi_A\xi_Z\rangle_{st}\left(1-e^{-u|t_1-t_2|}\right) \right],
\end{equation}
with $u\equiv\widetilde{\gamma}+\widetilde{D}-C'(\phi_{A,st})$ and $\phi_{A,st}$ the solution of $C(\phi_A)=(\widetilde{\gamma}+\widetilde{D})\phi_A$.
The derivative that appears in (\ref{correlationnz}) is:
\begin{equation}
 \frac{d^2\langle n_Z(t_1)n_Z(t_2)\rangle_\textrm{st}}{dt_1dt_2}=\Omega^2D^2\phi_{A,st}^2+\Omega\left[\widetilde{D}u\langle\xi_A\xi_Z\rangle_{st}e^{-u|t_1-t_2|}+\widetilde{D}\phi_{A,st}\delta(t_1-t_2)\right],
\end{equation}
with $\langle\xi_A\xi_Z\rangle_{st}=\widetilde{D}\phi_{A,st}\frac{2C'-(\gamma'+D')\phi_{A,st}}{2u^2}$.
Putting all the pieces together, one finally obtains:
\begin{eqnarray}
 K_\textrm{st}[n_I](t)&=&\langle n_I(t_0)n_I(t_0+t)\rangle_\textrm{st}-\langle n_I\rangle_\textrm{st}^2\\
&=&\Omega\widetilde{D}\phi_{A,st}\int_0^\infty dt'\hat{F}(t+t')+\Omega\widetilde{D}u\langle\xi_A\xi_Z\rangle_\textrm{st}\int_0^\infty ds\int_0^\infty dr\hat{F}(s)\hat{F}(r)e^{-u|t+s-r|}.\label{proteincorr}\nonumber
\end{eqnarray}
Proceeding in a similar way, one can derive:
\begin{eqnarray}
 K_\textrm{st}[n_A,n_I](t)&=&\Omega\langle\xi_A\xi_Z\rangle_\textrm{st}u\int_{0}^{\infty}dt'e^{-u(t+t')}\hat{F}(t')\\
K_\textrm{st}[n_I,n_A](t)&=&\Omega\langle\xi_A\xi_Z\rangle_\textrm{st}u\int_{0}^{\infty}dt'e^{-ut'}\hat{F}(t+t')+\Omega\widetilde{D}\langle\xi_A^2\rangle_\textrm{st}\int_{0}^{t}dt'e^{-ut'}\hat{F}(t-t'),\\
K_\textrm{st}[n_A](t)&=&\Omega\langle\xi_A^2\rangle_\textrm{st}e^{-ut},
\end{eqnarray}
with $K_\textrm{st}[n_u,n_v](t)\equiv\langle n_u(t_0+t)n_v(t_0)\rangle_\textrm{st}-\langle n_u\rangle_\textrm{st}\langle n_v\rangle_\textrm{st}$. This finally allows to express the correlation function for the total number of particles, $n=n_A+n_I$, as:
\begin{equation}
K_\textrm{st}[n](t)=K_\textrm{st}[n_I](t)+K_\textrm{st}[n_A,n_I](t)+K_\textrm{st}[n_I,n_A](t)+K_\textrm{st}[n_A](t).\label{proteincorrtot} 
\end{equation}
In this case, the average of $n$ again depends only on the average delay, $\langle n\rangle_\textrm{st}=\Omega\phi_A(1+D\langle\tau\rangle)$, but the second moment depends on the delay distribution in a more complicated way, through factors involving the integral of $\hat{F}(t)$.
 
\begin{figure}[h]
\centering
\includegraphics[scale=0.5,angle=0,clip]{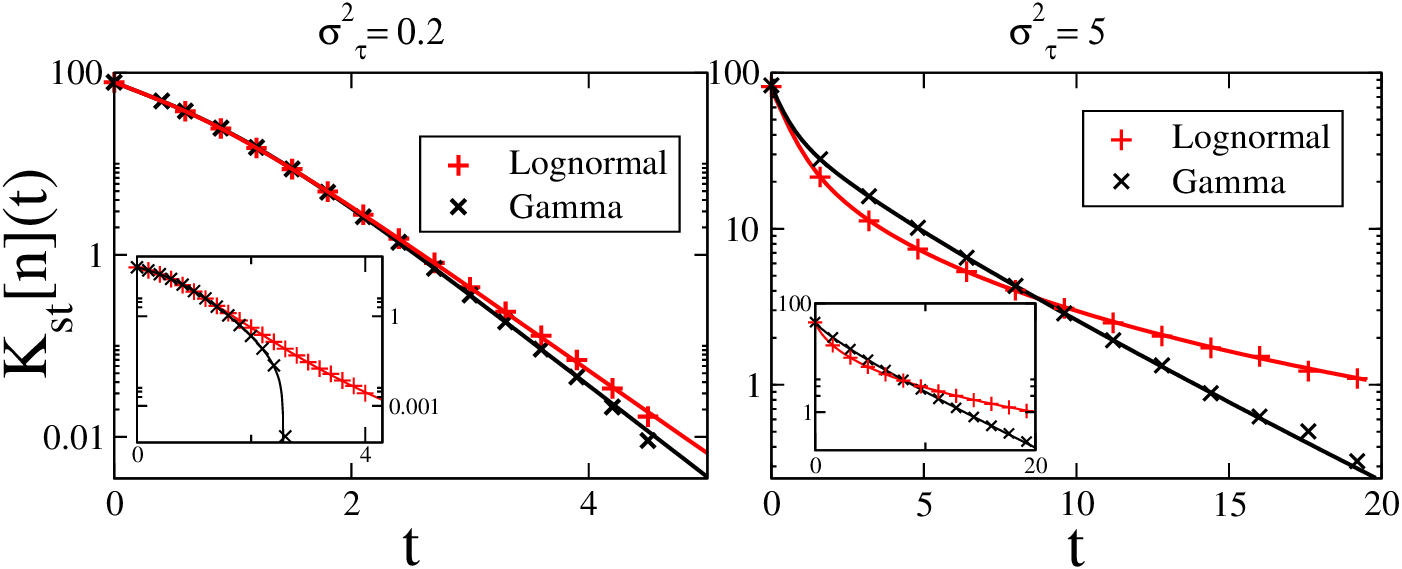}
\caption{Steady state correlation function for the total number of particles as a function of time, plotted in logarithmic scale, for two different types of delay distribution, gamma and lognormal, for two values of the variance of the delay: 0.2 (left panel) and 5 (right panel); in both cases the average delay is $\langle\tau\rangle=1$. The insets show the time correlation for the number of "infected" particles, $X_I$, which gives the largest contribution to the difference between different distributions. Symbols come from numerical simulations and lines from the theoretical analysis Eqs.(\ref{proteincorr}-\ref{proteincorrtot}). The creation rate is $C(n_A)=\frac{c\Omega}{1+\left(\epsilon\frac{n_A}{\Omega}\right)^2}$, parameters values are: $\Omega=100, c=1, \epsilon=0.4$ and $D=\gamma=1$. (Online version in colour.)
 \label{figure:correlationprot}} 
\end{figure}

In figure (\ref{figure:correlationprot}) this result is compared with numerical simulations, showing a very good agreement. Note that the treatment of the delayed reactions is exact, the only approximation coming from the use of van Kampen's expansion, which is needed when non-linearities are present, but whose error scales as $\Omega^{-1/2}$. Like in the previous case, the process in which the distribution of delay has fatter tail shows slower decay for the correlation function. 
\subsection{Full feedback}\label{sec:fullfeedback}
We now consider the case in which the creation rate depends on all present particles
\begin{equation}
 \emptyset {{C(n) \atop \longrightarrow}\atop{}} X, \,X {{\atop \Longrightarrow}\atop{\tau}} \emptyset,
\end{equation}
with $n$ the total (inert+active) number of $X$-particles. As noted before, this single-variable model can account for instantaneous plus delayed degradation, in the case that the degradation and ``contagion" rates, $\gamma$ and $D$ before, do not depend on the state of the system. For simplicity, we restrict our attention to this case. 
This process can be treated with the approach of the previous subsection introducing the additional variable $Z$,
\begin{equation}
 \emptyset {{C(n) \atop \longrightarrow}\atop{}} X+Z, \,X {{\atop \Longrightarrow}\atop{\tau}} \emptyset,
\end{equation}
with $n_Z(t)$ the corresponding random variable giving the number of $Z$ particles. 
We see that:
\begin{equation}
 n(t)=\int_{-\infty}^{t}dt'\frac{dn_Z(t')}{dt'}s(t',t)\label{nfeedback},
\end{equation}
with $s(t',t)$ the same as in the previous section. The probability distribution for $\{n,n_Z\}$ follows a master equation of the form:
\begin{equation}
 \frac{dP(n_Z,n,t)}{dt}=(E^{-1}E_Z^{-1}-1) C(n) P(n_Z,n,t)+(E-1)g(n_{Z},n)P(n_Z,n,t).\label{MEZ}
\end{equation}
Details of the derivation of the master equation in systems with delay are given in the appendix. Here, $g(n_Z,n)=\int_{0}^{\infty}dt'\langle C(n(t-t'))|n_Z(t),n(t)\rangle f(t')$, with $f(t)$ the probability density of the delay distribution, although, since we are only interested in the properties of variable $n$, we will not be using this expression. 
The key step in this case is to note that Eq.(\ref{MEZ}) allows us to derive the statistical properties (moments and correlations) of $n_Z(t)$ as a function of those of $n(t)$. Then, using (\ref{nfeedback}) we will be able to self-consistently derive the properties of $n$. More specifically, the approach proceeds as follows:\newline
Summing Eq.(\ref{MEZ}) over $n$, we can obtain an equation for the evolution of $P(n_Z,t)$, but that still depends on $n$ (in this step the contribution of the second term in Eq.(\ref{MEZ}) vanishes):
\begin{equation}
 \frac{dP(n_Z,t)}{dt}=(E_Z^{-1}-1) \sum_nC(n) P(n_Z,n,t)=(E_Z^{-1}-1)\langle C(n(t))|n_Z,t\rangle P(n_Z,t).\label{MEZ2}
\end{equation}
The two times probability distribution $P(n_{Z_1},t_1;n_{Z_2},t_2)$ follows a similar equation. Conditioning carefully, summing over the variable $n$ and considering separately the case $t_1=t_2$ (which turns out to be singular), we find:
\begin{eqnarray}
\frac{d^2P(n_{Z_1},t_1;n_{Z_2},t_2)}{dt_1dt_2}&=&(E_{Z_1}^{-1}-1)(E_{Z_2}^{-1}-1)\langle C(n_A(t_1))C(n_A(t_2))|n_{Z_1},t_1,n_{Z_2},t_2\rangle\times \nonumber\\
&&P(n_{Z_1},t_1;n_{Z_2},t_2)\label{MEZ2t}+\\
&&\delta(t_1-t_2)\left[(1-E_{Z_2})\delta_{n_{Z_1},n_{Z_2}}E_{Z_1}^{-1}+(1-E_{Z_2}^{-1})\delta_{n_{Z_1},n_{Z_2}}\right]\langle C(n_A)|n_{Z_1}\rangle P(n_{Z_1},t_1)\nonumber
\end{eqnarray}
From (\ref{nfeedback}) we easily obtain:
\begin{eqnarray}
 \langle n(t)\rangle&=&\int_{-\infty}^{t}dt_1\frac{d\langle n_Z(t_1)\rangle}{dt_1}\langle s(t_1,t)\rangle\\
\langle n(t)n(t')\rangle&=&\int_{-\infty}^{t}dt_1\int_{-\infty}^{t'}dt_2\frac{d^2\langle n_Z(t_1)n_Z(t_2)\rangle}{dt_1dt_2}\langle s(t_1,t)s(t_2,t')\rangle\label{corrfeedback}.
\end{eqnarray}
While (\ref{MEZ2}, \ref{MEZ2t}) imply:
\begin{eqnarray}
 \frac{d\langle n_Z\rangle}{dt}&=&\langle C(n(t))\rangle,\\
\frac{d^2\langle n_Z(t_1)n_Z(t_2)\rangle}{dt_1dt_2}&=&\langle C(n(t_1))C(n(t_2))\rangle+\delta(t_1-t_2)\langle C(n(t_1))\rangle
\end{eqnarray}
And we finally obtain the following set of integral equations for the moments:
\begin{eqnarray}
 \langle n(t)\rangle&=&\int_{-\infty}^{t}dt_1\langle C(n(t_1))\rangle\hat{F}(t-t_1)\\
\langle n(t)n(t')\rangle&=&\int_{-\infty}^{t}dt_1\int_{-\infty}^{t'}dt_2\langle C(n(t_1))C(n(t_2))\rangle\hat{F}(t-t_1)\hat{F}(t'-t_2)\nonumber\\
&&+\int_{-\infty}^{t}dt_1\langle C(n(t_1))\rangle\hat{F}(\max\{t,t'\}-t_1),
\end{eqnarray}
In the case of linear feedback, $C(n)=a+bn$, this system of equations is closed. For non-linear systems, one can use van Kampen's expansion as explained above.
In the steady state, one finds:
\begin{eqnarray}
 \langle n\rangle_{st}&=&\Omega\phi_\textrm{st},\hspace{1cm} \phi_\textrm{st}=C(\phi_\textrm{st})\langle \tau\rangle\label{averagefullfeedback}\\
K_\textrm{st}[n](t)&=&\int_{0}^{\infty}dx\left[\int_{0}^{t+x}dyK_\textrm{st}[n](t+x-y)\hat{F}(x)\hat{F}(y)+\int_{t+x}^{\infty}dyK_\textrm{st}[n](t+x-y)\hat{F}(x)\hat{F}(y)\right]\nonumber\\
&+&\Omega C(\phi_\textrm{st})\int_{0}^{\infty}dx\hat{F}(t+x)\label{corfullfeedback}
\end{eqnarray}

Eq. (\ref{averagefullfeedback}) shows that the steady state number of particles depends only on the average delay. Eq. (\ref{corfullfeedback}) shows that the correlations depend on the delay distribution in a non-trivial way. The analysis of this equation is left for future work.

\section{Delayed creation}\label{dc}
We now turn our attention to the case in which the creation reaction, that is initiated stochastically, takes a finite time to be completed. For simplicity we assume that the degradation reaction is instantaneous. Schematically, we have:
\begin{eqnarray}
\label{eq:processcreation}
\emptyset {{C(n) \atop \longrightarrow}\atop{}}{{ {}\atop \Longrightarrow}\atop{\tau}}X,\hspace{0.5cm} X {{\gamma \atop \longrightarrow}\atop{}} \emptyset.
\end{eqnarray}
In this case, if the creation rate does not depend on the number of particles, $n$, then the delay in the creation is completely irrelevant, since the probability that a new particle appears at time $t$ is equal to the probability that its creation started at a time $t-\tau$, but this equal to the probability that a particle starts its creation at time $t$ (with a shift in the time if $C$ is time-dependent), so the process is completely equal to one with instantaneous creation.

Following \cite{creationdelay} we will adopt here an approach different from that of the previous sections, that, besides the moments, will allow us to obtain an expression for the full probability distribution. For completeness we will explain here the method in some detail. For additional considerations, the reader is referred to \cite{creationdelay}. 
In the appenix it is shown that the master equation of the process (\ref{eq:processcreation}) is:
\begin{eqnarray}
 \frac{\partial P(n,t)}{\partial t}&=&(E-1)[\gamma nP(n,t)]+(E^{-1}-1)\left[\sum_{n'=0}^{\infty}\int_0^\infty d\tau C(n')P(n',t-\tau;n,t)f(\tau)\right]\label{master_delaydistrib}.
\end{eqnarray}
The master equation (\ref{master_delaydistrib}) can be written as:
 \begin{equation}
\label{master_effdelay}
 \frac{\partial P(n,t)}{\partial t}=(E-1)[\gamma nP(n,t)]+(E^{-1}-1)[\tilde C(n,t)P(n,t)],
\end{equation}
where the effective creation rate, $\tilde{C}(n,t)$, is given by:
\begin{equation}\label{eq:ctilde}
\tilde C(n,t)=\int_0^\infty d\tau f(\tau)\langle C(n'(t-\tau))|n(t)\rangle.
\end{equation}
The conditional probability $P(n,t|n_0,t_0)$ follows a master equation identical to (\ref{master_delaydistrib}) with all the probabilities conditioned to $n_0$ at time $t_0$. From it, and using that $\langle n(t)|n(t_0)\rangle=\sum_nnP(n,t|n(t_0),t_0)$, we obtain the following evolution equation for the conditional average:
\begin{eqnarray}
\frac{d\langle n(t)|n(t_0)\rangle}{dt}&=&-\gamma\langle n(t)|n(t_0)\rangle+\int_0^\infty d\tau f(\tau)\langle C(n(t-\tau))|n(t_0)\rangle,\label{averagedelaygen}
\end{eqnarray}
for $t\ge 0$, with initial condition $\langle n(t_0)|n(t_0)\rangle= n(t_0)$.

The knowledge of the steady value $\tilde C_\textrm{st}(n)\equiv \lim_{t\to\infty}\langle C(n'(t-\tau))|n,t)\rangle=\langle C(n'),-\tau|n\rangle_\textrm{st}$, allows the calculation of the steady-state probabilities $P_{st}(n)$, obtained by imposing $\frac{\partial P(n,t)}{\partial t}=0$ in Eq.(\ref{master_effdelay}),  as \cite{VK}:
\begin{equation}\label{probgeneral}
 P_\textrm{st}(n)=P_\textrm{st}(0)\prod_{k=0}^{n-1}\frac{\tilde C_\textrm{st}(k)}{\gamma(k+1)}=\frac{P_\textrm{st}(0)}{\gamma^nn!}\prod_{k=0}^{n-1}\tilde C_{st}(k),
\end{equation}
$P_\textrm{st}(0)$ is fixed by the normalization condition. All is left to do now is to compute the effective creation rate $\tilde C_\textrm{st}(n)$.

The effective creation rate will be computed using expression (\ref{averagedelaygen}). In the general case of nonlinear creation rate, we will use van Kampen's expansion to linearize $C(n)$ around the macroscopic component of $n$. We have: $C(n)=\Omega C(\phi)+\Omega^{1/2}C'(\phi)\xi$, so 
\begin{equation}
 \langle C(n'(t-\tau))|n(t)\rangle=\Omega C(\phi(t-\tau))+\Omega^{1/2}C'(\phi(t-\tau))\langle\xi'(t-\tau)|\xi(t)\rangle\label{cexpanded}
\end{equation}
using (\ref{averagedelaygen}) we obtain:
\begin{eqnarray}
\frac{d\phi(t)}{dt}&=&-\gamma\phi(t)+\int_0^\infty d\tau f(\tau)C\left(\phi(t-\tau)\right),\label{vkfluc2}\\
\frac{d\langle\xi(t')|\xi(t)\rangle}{dt'}&=&-\gamma\langle\xi(t')|\xi(t)\rangle+\int_0^{\infty} d\tau f(\tau)C'\left(\phi(t-\tau)\right)\langle\xi(t')-\tau|\xi(t)\rangle\label{vkflucgen}
\end{eqnarray}
Equation (\ref{vkfluc2}) is in general a non-linear integro-differential equation, that can be difficult to solve. Here, however, we will focus on the cases in which (\ref{vkfluc2}) has a stable steady state as a single attractor, which is the solution of $\gamma\phi=C(\phi)$. This is the regimen in which the validity of van Kampen's expansion is guaranteed. 

We reach now a delicate point. Eq.(\ref{vkflucgen}) is a (linear) integro-differential equation. To solve it, we would need an initial condition in the whole interval $(-\infty,t)$ but we only know a one-time condition $\langle\xi(t'=t)|\xi(t)\rangle=\xi(t)$. We will circumvent this difficulty by assuming that, over the steady state, the system is statistically invariant under time-inversion, which implies $\langle\xi(t+t_1)|\xi(t)\rangle=\langle\xi(t-t_1)|\xi(t)\rangle$. This condition, together with the value of $\xi$ at time $t$, allows to find the solution of (\ref{vkflucgen}).
The time-reversal invariance assumption in the steady state is fulfilled by any Markovian system that follows detailed balance. Our system follows detailed balance (as any one-step process \cite{VK}), but, due to the presence of delay, it is not Markovian. So the time-reversal invariance is an assumption, whose validity needs to be checked. It was shown in \cite{creationdelay} that in this system the assumption is approximately valid.

\begin{figure}[h]
\centering
\includegraphics[scale=0.25,angle=0,clip]{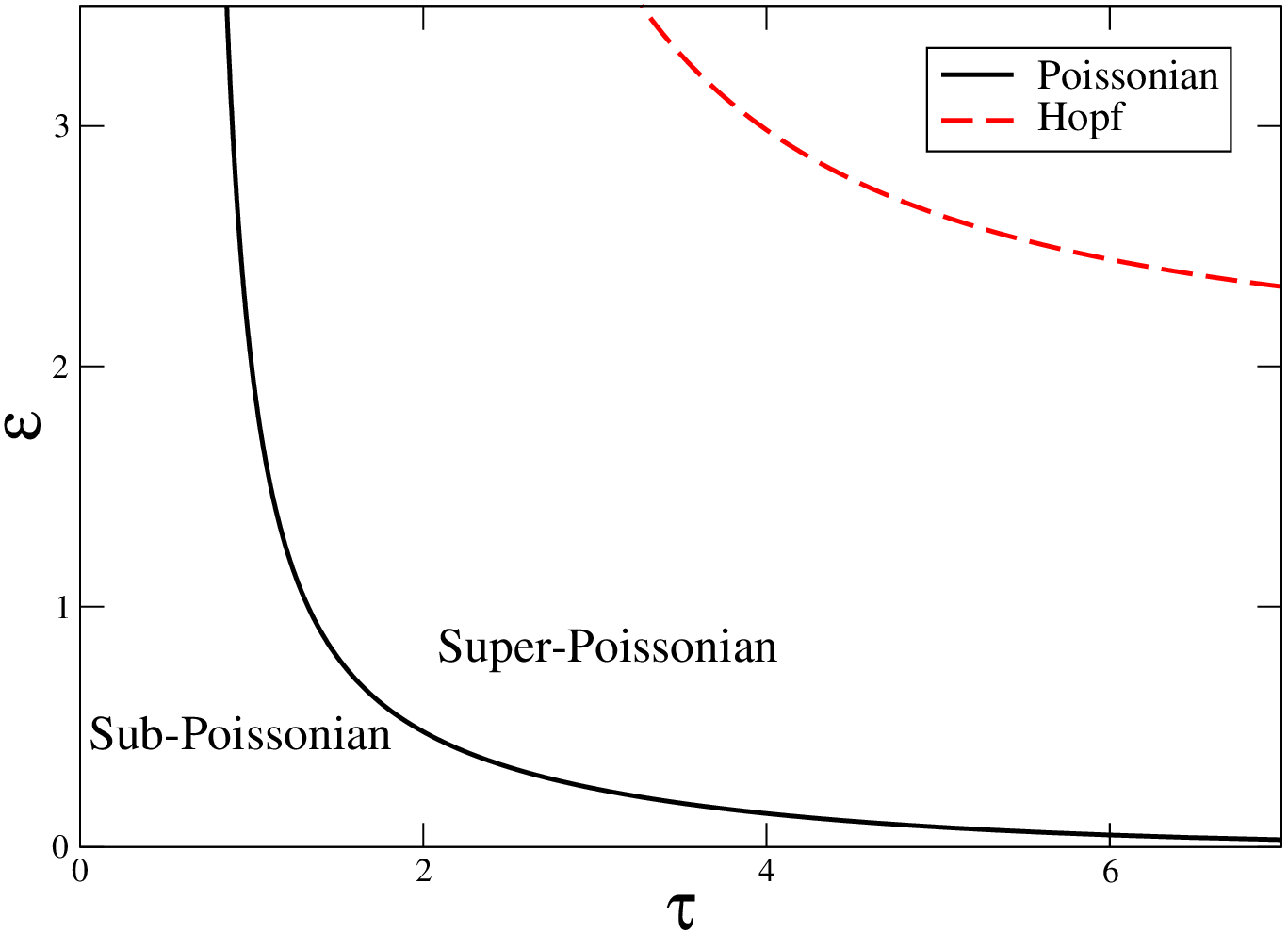}
\caption{Relative size of the variance respect mean value for the number of particles, in the $\tau-\epsilon$ plane, for creation with constant delay and a negative feedback given by a creation rate of the form $\frac{c}{1+\left(\epsilon\phi\right)^2}$ (note that $\epsilon$ is the strength of the negative feedback). The "Poissonian line", $\sigma^2[n]=\langle n\rangle$, obtained through the approximation (\ref{vksigma}), marks the transition from sub-Poissonian to Super-Poissonian fluctuations, while the Hopf line marks the Hopft transition into oscillatory behavior in the deterministic system. Parameters values are: $c=1,$ $D=\gamma=1$. (Online version in colour.)
 \label{figure:phasespace}}
\end{figure}

In the case of constant delay, $f(t)=\delta(t-\tau)$, the time-reversal symmetric solution of (\ref{vkflucgen}) is \cite{bratsun,creationdelay}:
\begin{eqnarray}
\langle\xi(t+t_1)|\xi(t)\rangle&=&\xi(t)h(t_1)\label{xiresult}\\
h(t_1)&\equiv&\begin{cases}\frac{e^{-\lambda t_1}-\zeta e^{\lambda (t_1-\tau)}}{1-\zeta e^{-\lambda\tau}},&\text{if }0\le t_1\le\tau\\ \\
e^{-\gamma (t_1-k\tau)}h(k\tau)-\alpha\int_{k\tau}^{t_1}dt'\,h(t'-\tau)e^{\gamma (t'-t_1)},&\text{if }k\tau \le t_1\le(k+1)\tau, k=1,2,\cdots
\end{cases}\\
\lambda&\equiv&\sqrt{\gamma^2-\alpha^2}, \hspace{1cm}\zeta\equiv\frac{\gamma-\lambda}{\alpha},\hspace{1cm}\alpha\equiv-C'(\phi_st)\nonumber
\end{eqnarray}
and using (\ref{cexpanded}) we finally obtain:
\begin{equation}
 \tilde C(n)=\Omega\left[C(\phi_{st})-\phi_\textrm{st}C'(\phi_\textrm{st})h(\tau)\right]+C'(\phi_\textrm{st})h(\tau)n,
\end{equation}
where $\phi_{st}$ is the steady state solution of (\ref{vkfluc2}).
From Eq.(\ref{probgeneral}) one can obtain the steady-state probabilities $P_{st}(n)$. The mean value and variance are given by:
\begin{eqnarray}
 \langle n\rangle_\textrm{st}&=&\Omega\phi_{st}\label{vkaverage}\\
\sigma^2_\textrm{st}&=& \frac{\langle n\rangle_\textrm{st}}{1-\gamma^{-1}\varPhi'(\phi_{st})h(\tau)}\label{vksigma}.
\end{eqnarray}
From (\ref{vksigma}) one can see that, interestingly, in the case of negative feedback ($C'(\phi_{st})<0$), as the delay is increased the fluctuations change from sub-Poissonian ($\sigma^2<\langle n\rangle$) to super-Poissonian ($\sigma^2>\langle n\rangle$). This is illustrated in figure (\ref{figure:phasespace}), where we also show the line of the Hopf bifurcation for the deterministic system. It is usually obtained that a negative feedback reduces the magnitude of the fluctuations \cite{oudenardenpnas}, when delay present we see that this negative feedback can change totally its effect, giving rise to an increase of the fluctuations.

The time correlation function can also be obtained from (\ref{xiresult}), as:
\begin{equation}
 K_\textrm{st}[n](t)=\sigma^2_\textrm{st}h(t)
\end{equation}
In the case of negative feedback it becomes non-monotonic, developing peaks of alternating sign at approximately multiples of the delay, signaling the presence of stochastic oscillations.
For positive feedback, the time correlation is always positive, but not necessarily monotonic.

We will finish by noting that the ``effective Markovian reduction'' method used in the previous section can also be used for the case of delay in the creation with feedback. To be completely general, we allow two delays, one in the creation (with probability density $f_c(t)$), and one in the degradation (with probability density $f_d(t)$). The process is schematized as follows:
\begin{eqnarray}
\emptyset{{C(n) \atop \longrightarrow}\atop{}}{{\atop \Longrightarrow}\atop{\tau_c}}X,\hspace{0.5cm} X {{ {}\atop \Longrightarrow}\atop{\tau_d}}\emptyset,
\end{eqnarray}
with $\tau_{c/d}$ random variables distributed according to $f_{c/d}(t)$. With the addition of two new variables, the process can be rewritten as:
\begin{eqnarray}
\emptyset {{C(n) \atop \longrightarrow}\atop{}} Z+Y,\hspace{0.5cm} Y{{ {}\atop \Longrightarrow}\atop{\tau_c}}X,\hspace{0.5cm} X {{ {}\atop \Longrightarrow}\atop{\tau_d}}\emptyset,
\end{eqnarray}
which allows us to note that:
\begin{equation}
 n(t)=\int_{-\infty}^{t}dt'\frac{dn_Z(t')}{dt'}\tilde{s}(t',t).\label{nt}
\end{equation}
In this case, the survival function $\tilde{s}(t',t)$ is defined as: $\tilde s(t',t)=1$, if $t\in(t'+\tau_c(t'),t'+\tau_c(t')+\tau_d(t'))$ , and $\tilde s(t',t)=0$, otherwise, being $\tau_c(t')$ and $\tau_d(t')$ random times obtained from the corresponding pdf's $f_c(\tau_c)$ and $f_d(\tau_d)$. $\tilde{s}(t',t)$ is equal to one if a virtual particle that initiated its creation at time $t'$ finished it at some intermediate time $t''<t$ and since then had a lifetime greater that $t-t''$, so that it is still alive at $t$, being zero otherwise. It follows that:\begin{eqnarray}
\langle \tilde{s}(t_1,t)\rangle&=&\int_{0}^{t-t_1}dt'f_c(t')\hat F_d(t-t_1-t')\\
\langle \tilde{s}(t_1,t)\tilde{s}(t_2,t')\rangle&=&\begin{cases}\langle \tilde s(t_1,t)\rangle\langle \tilde s(t_2,t')\rangle,&\text{if }t_1\ne t_2,\\
\int_{0}^{\min\{t,t'\}-t_1}dt''f_c(t'')\hat F_d(\max\{t,t'\}-t_1-t''),&\text{if } t_1=t_2.
\end{cases}
\end{eqnarray}
In the case that the creation rate $C(n)$ does not depend on the number of $X$-particles, the number of $Z$-particles follows a Markovian process (Poisson process), and the properties of $n$ can be derived from (\ref{nt}). If the creation rate depends on the number of $X$-particles i.e. if feedback is present, the properties of $n_Z$ can be derived formally as a function of $n$ and then the properties of $n$ can be derived self-consistently trough (\ref{nt}), as done in subsection (\ref{sec:fullfeedback}).

\section{Comments and conclusions}\label{comments}
In this paper we have analyzed general stochastic birth and death models that include delay. We have presented three different methods that together constitute a general toolbox to study stochastic models including delay. 

In sub-section (\ref{sec:simple}) we have shown that when the creation rate is independent of the state of the system (no feedback) and the initiation of the delayed degradation and the instantaneous degradation are first order reactions (rate not depending on the state of the system), the process can be solved fully in an exact fashion for general distributions of delay, showing always Poissonian character and a monotonically decreasing time correlation function given by (\ref{corrsimple}).
 
In sub-sections (\ref{sec:elaborated}), (\ref{sec:fullfeedback}) we have considered a more general process with delay in the degradation step, allowing the initiation of the delay degradation and the instantaneous degradation to be higher order reactions, as well as the presence of feedback in the creation rate. The method allows to reduce the system to a Markovian one, where usual techniques can be used. Explicit expressions for the time correlation for general delay distributions were obtained. In this case the correlation might be non-monotonic, if feedback is present, but typically decreases monotonically.

Section (\ref{dc}) shows that when the delay appears in the creation reaction and feedback is present, the delay typically has more dramatic consequences. In the case of fixed delay, it is shown that for negative feedback, the fluctuations are amplified as the delay increases, going beyond the level found when no feedback is present, and the time correlation function becomes oscillatory, alternating positive and negative values at approximately multiples of the delay. In the positive feedback case, again for fixed delay, the fluctuations are reduced with increased delay and the time correlation function remains always positive.

\section{Appendix: Calculation of $P(n,t)$ in the simple case of delayed degradation}
We start by considering the case $n=0$. For the sake of simplicity, we focus on the case with creation rate, $C$, independent of time, but the generalization to time-dependent $C$ is straightforward. Since the time origin is taken at $t=0$, the probability of observing zero particles at time $t>0$ is equal to the following limit:
\begin{equation}
P(0,t)=\lim_{M\rightarrow\infty}\prod_{i=0}^{M-1}\left[1-C\Delta t+C\Delta t F(t-t_i)+o(\Delta t)\right],\label{P0tini}
\end{equation}
with $\Delta t\equiv \frac{t}{M}$ playing the role of a small time-increment and $t_i\equiv i\Delta t$. This expression follows from the fact that, in order to find the system with zero particles at time $t$, in every previous infinitesimal time interval ($t'\in[t_i,t_{i+1}), i=0,\dots,M-1$) one of the following two (incompatible) events must take place: either a particle is not created (probability $1-C\Delta t$) or a particle is created with a lifetime smaller that $t-t_i$ (probability $C\Delta t F(t-t_i)$). We now have:
\begin{equation}
 \log P(0,t)=\lim_{M\rightarrow\infty}\sum_{i=0}^{M-1}\left[-C\hat{F}(t-t_i)+o(\Delta t)\right]\Delta t=-C\int_{0}^tdt'\hat{F}(t-t'),
\end{equation}
with $\hat{F}(t)\equiv1-F(t)$, so we find
\begin{equation}
P(0,t)=e^{-C\int_{0}^tdt'\hat{F}(t-t')}.\label{P0t}
\end{equation}

Following a similar line of reasoning, $P(n,t)$ can be computed as:
\begin{equation}
 P(n,t)=\lim_{M\rightarrow\infty}\sum_{i_1=0}^{M-1}\sum_{i_2=i_1+1}^{M-1}\cdots\sum_{i_n=i_{n-1}+1}^{M-1}\
 \prod_{l=1}^n\left[C\Delta t\hat{F}\left(t-t_{i_l}\right)\right]\prod_{{0\le j\le M-1}\atop{j\ne i_1,i_2,\dots,i_n}}\left[1-C\Delta t\hat{F}(t-t_{i_j})\right]
\end{equation}
This expression results from the consideration of choosing the times $(t_{i_1},\dots,t_{i_n})$ at which the $n$ particles are created and survive up to $t$. The $l$-th particle is  created with probability $C\Delta t$ and survives up to $t$ with probability $\hat{F}\left(t-t_{i_l}\right)$. The other factor comes from the fact that at the other time intervals either a particle is not created or it is created but dies before $t$.

Using 
\begin{equation}
 \lim_{M\rightarrow\infty}\prod_{{0\le j\le M-1}\atop{j\ne i_1,i_2,\dots,i_n}}\left[1-C\Delta t\hat{F}(t-t_{i_j})\right]=e^{-C\int_0^tdt'F(t-t')}
\end{equation}
and replacing the sums by integrals in the limit $M\to \infty$
\begin{equation}
\int_0^tdt_1C\hat F(t-t_1)\int_{t_1}^t dt_2C\hat F(t-t_2)\dots\int_{t_{n-1}}^t dt_n C\hat F(t-t_n)=\frac{C^n}{n!}\left[\int_0^t dt' F(t-t)\right]^n
\end{equation}
we finally obtain:
\begin{equation}
P(n,t)=e^{-C\int_{0}^tdt'\hat{F}(t-t')}\frac{C^n\left[\int_{0}^tdt'\hat{F}(t-t')\right]^n}{n!},
\end{equation}
that is, a Poisson distribution with average $\langle n(t)\rangle=C\int_{0}^tdt'\hat{F}(t-t')$. In the steady state (found as the limit $t\rightarrow\infty$), the average becomes $\langle n(t)\rangle=C\langle \tau\rangle$. Remarkably, this Poissonian character is completely independent of the form of the delay distribution. As commented above, this result can be easily generalized to the case in which the creation rate depends on time, $C\rightarrow C(t)$, obtaining again a Poisson distribution with average $\int_{0}^tdt'C(t')\hat{F}(t-t')$.

\section{Appendix: derivation of the master equation in a system with delay}
Here we derive the master equation of the process (\ref{eq:processcreation}). We consider first the case of fixed delay $\tau$. We start with the following identity:
\begin{equation}
 P(n,t+\Delta)=\sum_{n'}P(n,t+\Delta;n',t)=P(n,t+\Delta;n+1,t)+P(n,t+\Delta;n-1,t)+P(n,t+\Delta;n,t)+o(\Delta).\label{identitycreation}
\end{equation}
It is immediate to see that $P(n,t+\Delta;n+1,t)=\gamma(n+1)\Delta P(n+1,t)$. In the case of fixed delay, the second sum can be evaluated introducing a tree-times probability as:
\begin{eqnarray}
 P(n,t+\Delta;n-1,t)&=&\sum_{n'}P(n,t+\Delta;n-1,t;n',t-\tau)\nonumber\\
&=&\sum_{n'}P(n,t+\Delta|n-1,t;n',t-\tau)P(n',t-\tau;n-1,t).
\end{eqnarray}
Now, $P(n,t+\Delta|n-1,t;n',t-\tau)=C(n')\Delta+o(\Delta)$. Expanding in a similar way the term $P(n,t+\Delta;n,t)$, and taking the limit $\Delta\rightarrow0$, we can obtain the master equation of the process:
\begin{eqnarray}
 \frac{\partial P(n,t)}{\partial t}&=&(E-1)[\gamma nP(n,t)]+(E^{-1}-1)\left[\sum_{n'=0}^{\infty}C(n')P(n',t-\tau;n,t)\right]\label{master_delay}.
\end{eqnarray}
In the case of distributed delay, we start considering a discrete distribution of delays i.e. $\tau=\tau_1,\dots,\tau_M$ with corresponding probabilities $f(\tau_1),\dots f(\tau_M)$. The continuum limit can then be obtained making $M\rightarrow\infty$. The creation term in (\ref{identitycreation}) can be written as:
\begin{eqnarray}
&&P(n,t+\Delta;n-1,t)=\sum_{n_1,\dots n_M}P(n,t+\Delta;n-1,t;n_1,t-\tau_1;\dots;n_M,t-\tau_M)\\
&&=\sum_{n_1,\dots,n_M}P(n,t+\Delta|n-1,t;n_1,t-\tau_1;\dots;n_M,\tau_m)P(n_1,t-\tau_1;\dots;n_M,t-\tau_M;n-1,t)\nonumber.
\end{eqnarray}
Now, $P(n,t+\Delta|n-1,t;n_1,t-\tau_1,\dots,n_M,\tau_m)=\sum_{i=1}^MC(n_i)f(\tau_i)\Delta+o(\Delta)$, that is, the probability that a particle started its creation at time $t-\tau_i$ with a creation time equal to $\tau_i$. Replacing in the previous equation and performing the appropriate sums we obtain:
\begin{equation}
P(n,t+\Delta;n-1,t)=\sum_{n'}\sum_{i=1}^MC(n')f(\tau_i)P(n',t-\tau_i;n-1,t)\Delta+o(\Delta)
\end{equation}
that in the continuum limit reduces to $\sum_{n'}\int_0^\infty d\tau C(n')f(\tau)P(n',t-\tau;n-1,t)$.
Considering in a similar way the other terms in (\ref{identitycreation}) and taking the limit $\Delta\rightarrow 0$ one can obtain the master equation for distributed delay (\ref{master_delaydistrib}).

\section{Acknowledgements}
We acknowledge financial support by the MICINN (Spain) and FEDER (EU) through project FIS2007-60327. L.F.L. is supported by the JAE Predoc program of CSIC.

\bibliographystyle{plain}

\end{document}